\documentclass[letterpaper,11pt]{article}

\usepackage{fancyhdr}
\usepackage{amsmath}
\usepackage{amsbsy}
\usepackage{amssymb}
\usepackage{amsthm}
\usepackage{amscd}
\usepackage{amsfonts}
\usepackage{supertabular}
\usepackage{graphics}
\usepackage{graphicx}
\usepackage{verbatim}
\usepackage{subfigure}
\usepackage{epsfig}
\usepackage{xspace}
\usepackage{euscript}
\usepackage{alltt}
\usepackage{boxedminipage}
\usepackage{float}
\usepackage{times}
\usepackage[colorlinks]{hyperref}
\usepackage[square,numbers,compress]{natbib}
\usepackage{setspace}
\usepackage{epstopdf}
\usepackage{authblk}

\usepackage{algorithm}
\usepackage{algorithmic}

\newcommand{\qvec}[1]{{\boldsymbol{#1}}}

\begin{document}

\title{Connecting discrete particle mechanics to continuum granular micromechanics: Anisotropic continuum properties under compaction}
\author[1]{Payam~Poorsolhjouy \thanks{E-mail address: ppoorsol@purdue.edu (P. Poorsolhjouy).}}
\author[1,2]{Marcial~Gonzalez \thanks{Corresponding author. Tel.: +1 765 494 0904; fax: +1 765 496 7537.
E-mail address: marcial-gonzalez@purdue.edu (M. Gonzalez).}}

\affil[1]{School of Mechanical Engineering, Purdue University, West Lafayette, IN 47907, USA}
\affil[2]{Ray W. Herrick Laboratories, Purdue University, West Lafayette, IN 47907, USA}

\maketitle
\begin{abstract}
A systematic and mechanistic connection between granular materials’ macroscopic and grain level behaviors is developed for monodisperse systems of spherical elastic particles under die compaction. The Granular Micromechanics Approach (GMA) with static assumption is used to derive the stiffness tensor of transversely isotropic materials, from the average behavior of particle-particle interactions in all different directions at the microscale. Two particle-scale directional density distribution functions, namely the directional distribution of a combined mechano-geometrical property and the directional distribution of a purely geometrical property, are proposed and parametrized by five independent parameters. Five independent components of the symmetrized tangent stiffness tensor are also determined from discrete particle mechanics (PMA) calculations of nine perturbations around points of the loading path. Finally, optimal values for these five GMA parameters were obtained by minimizing the error between PMA calculations and GMA closed-form predictions of stiffness tensor during the compaction process. The results show that GMA with static assumption is effective at capturing the anisotropic evolution of microstructure during loading, even without describing contacts independently but rather accounting for them in an average sense.
\end{abstract}


\section{Introduction}

The macroscopic, or global, mechanical behavior of materials is a direct function of their  microstructure and associated micromechanical characteristics \cite{Chang-Packing}. This is most clear when dealing with granular materials where the microstructure is composed of grains and, therefore, their arrangement clearly affects the macroscopic behavior. Modeling the behavior of these materials using macroscopic tensorial continuum mechanics results in an obvious neglect of the effects of the granular microstructure and its evolution, as well as of micro-mechanical phenomena taking place at grain scale, on the macroscopic mechanical response.

In order to incorporate microstructural properties of the material into its global behavior, many different schemes working in different spatial scales are available. In the broadest sense, all models can be categorized into two distinct groups, namely (i) {\it discrete models} where, according to the length scale being resolved, grains/particles/molecules/atoms are regarded as material's building blocks (e.g., atomic models \cite{dharmawardhana2014quantum,pellenq2009realistic}, molecular-dynamics \cite{hansson2002molecular, masoumi2017intermolecular},  bead-spring models \cite{milchev1996static}, dynamic discrete element methods \cite{zhu2008discrete,donze2009advances} and quasi-static particle mechanics approaches \cite{Gonzalez-2012,Gonzalez-2016, yohannes2016evolution, Gonzalez-2018, cusatis2011lattice}); (ii) {\it continuum models} where the material point is assumed to be a homogeneous continuum body whose behavior is interpreted in terms of tensorial quantities such as stress, strain, and stiffness \cite{lubliner1989plastic, miller2000continuum, chaboche2008review}.

Discrete models in principle can be used to derive highly accurate results with high fidelity. However, they rely upon correctly attainable details of material microstructure and of micro-mechanical phenomena. Continuum models, on the other hand, derive material response without exact consideration of microstructure and therefore, lack a connection between macroscopic observable behavior and its microscopic roots. The Granular Micromechanics Approach (GMA) provides a robust framework for connecting these two groups of models and bridges the gap between them. This is achieved by deriving such continuum macroscopic response from the study of average behavior of particle-particle interactions in all different directions at the microscale \cite{Ranga-2015,Viraj-2013,Payam-LoadPath,Payam-BandGap}. In doing so, GMA delivers the most crucial advantages of discrete models, i.e., it incorporates material's micromechanical features, microstructural effects, and load-path dependent anisotropic evolution, while avoiding the large computational cost associated with discrete models. It is worth noting that grain-pair interactions in GMA do not represent the behavior of two isolated grains, but rather, that of a grain-pair embedded in the granular microstructure. The global anisotropic continuum behavior of the granular material is then derived from the effective and directional behavior of grain-pair interactions. Therefore, the most critical element in deriving a predictive GMA model of any given material is formulating force-displacement relationships for grain-pair interactions.

In this communication, we address the issue of formulating particle-particle interactions, and their anisotropic evolution, during die compaction of a monodisperse system of spherical elastic particles. We focus on developing a systematic and mechanistic approach for identifying these relationships from discrete particle mechanics simulations of the granular system. The proposed methodology, therefore, effectively connects discrete particle mechanics to continuum granular micromechanics. Next, we briefly describe the GMA with static assumption adopted in this work.

\section{GMA with Static Assumption}

The GMA can take two general approaches, namely the method with a kinematic constraint and that with a static constraint. The approach with a kinematic constraint assumes that inter-particle displacements $\delta_i \in V$ can be derived as the projection of the macroscopic strain tensor $\epsilon_{ij} \in V^2$ on the particle-particle relative position $l_i \in V$, i.e., $\delta_i = \epsilon_{ij} l_j$. On the other hand, the approach with static constraint assumes a relationship between macroscopic stress tensor $\sigma_{ij} \in V^2$ and inter-particle force vectors $f_i \in V$. Here $V$ is a three-dimensional real vector space.

The GMA with static assumption enforces the kinematic constraint in a weak sense, that is
\begin{equation}
\epsilon_{pq}
=
\arg\min_{\epsilon_{pq} \in V^2}
\left\lVert \sum_{\alpha=1}^{N_c}
\left(\delta_i^\alpha-\epsilon_{ij} l_j^\alpha\right)
\right\lVert
\label{Residual}
\end{equation}
where $N_c$ denotes the total number of contacts $\alpha$ in the representative volume element. Therefore, the GMA with static constraint minimizes the sum over all contacts of the residual difference between the inter-particle displacement and the projection of macroscopic strain tensor on the vector joining the centroids of the particles forming each pair-contact $\alpha$. Furthermore, the Principle of Virtual Work (PVW) states the equality of macroscopic strain energy density and the volume average of inter-particle energies, that is
\begin{equation}
W
=
\sigma_{ij} \epsilon_{ij}
=
\frac{1}{V}\sum_{\alpha=1}^{N_c}
f_i^\alpha \delta_i^\alpha
\label{PVW}
\end{equation}
where $V$ is the volume of the representative volume element. By replacing \eqref{Residual} into \eqref{PVW}, the following relationship between the macroscopic stress tensor and the microscopic inter-particle forces is obtained
\begin{equation}
f_i=\sigma_{ij}N_{jq}^{-1}l_q;\ \ \ \ \ \ \ \text{with}\ \ \ \ N_{ij}=\frac{1}{V}\sum_{\alpha=1}^{N_c}l_i^\alpha l_j^\alpha
\label{StaticConst}
\end{equation}
where $N_{ij} \in V^2$ is the second rank fabric tensor. The above relationship is commonly known as the static constraint, and thus the name of the method. With some algebraic manipulation, the following expressions for the macroscopic strain and compliance $S_{ijkl} \in V^4$ tensors are obtained
\begin{subequations}
\begin{align}
S_{ijkl}
=
\frac{\partial \epsilon_{ij}}{\partial \sigma_{kl}}
=
N_{jp}^{-1}N_{lq}^{-1}\frac{1}{V}\sum_{\alpha=1}^{N_c}s_{ik}^\alpha l_p^\alpha l_q^\alpha
\\
\epsilon_{ij}
=
S_{ijkl} \sigma_{kl}
\end{align}
\label{epsilon-Sijkl}
\end{subequations}
where $s_{ij}^\alpha \in V^2$ is the local compliance tensor connecting inter-particle force and displacement of contact $\alpha$, that is $\delta_i^\alpha=s_{ij}^\alpha f_j^\alpha$. For a more detailed description of the above formulation see \cite{Payam-2d}. 

\begin{figure}[htbp]
	\centering
	\includegraphics[scale=0.27]{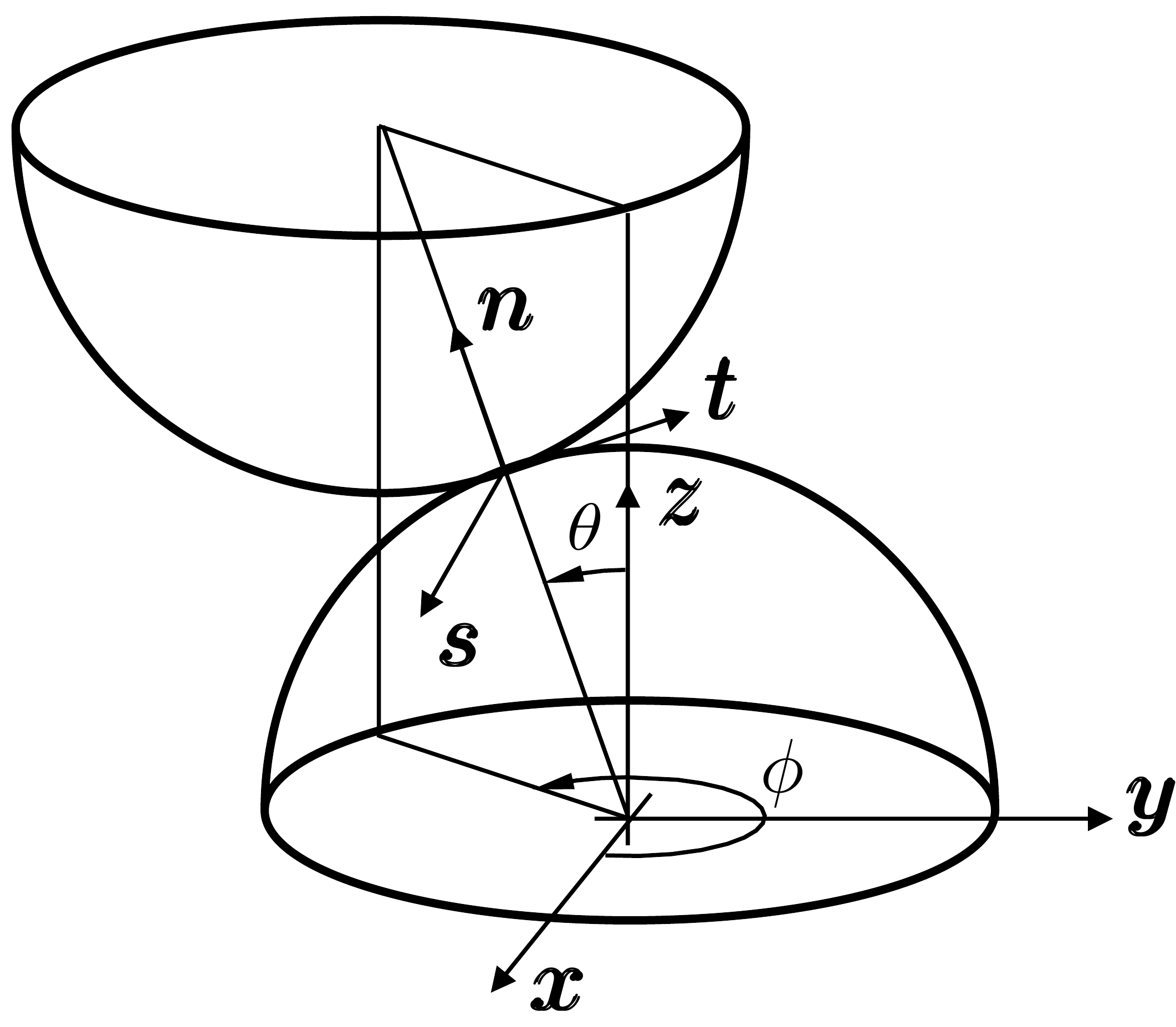}
	\caption{Two grains in contact and the local coordinate system along with the global Cartesian and spherical coordinate systems.}
	\label{fig-nst}
\end{figure}

For convenience, the inter-particle force-displacement relationship $f_i^\alpha(\delta_j^\alpha)$ can be formulated in a local coordinate system defined by the following three mutually orthogonal axes: one normal axis $n_i^\alpha$ in the direction of the vector joining the centroids of the  particles, and two tangential axes $s^\alpha_i$ and $t^\alpha_i$ (see Figure \ref{fig-nst}). Therefore, the microscopic constitutive relationship can be expressed on the local coordinate systems as follows
\begin{equation}
\begin{Bmatrix}
\delta^\alpha_n\\ \delta^\alpha_s\\\delta^\alpha_t
\end{Bmatrix}=\begin{bmatrix}
s^\alpha_n & 0 & 0\\0 & s^\alpha_s & 0\\ 0 & 0 & s^\alpha_t
\end{bmatrix}\begin{Bmatrix}
f^\alpha_n\\ f^\alpha_s\\f^\alpha_t
\end{Bmatrix}
\end{equation}
where the local compliance tensor is assumed to be symmetric by neglecting cross-coupling terms and thus $s^\alpha_n=1/k^\alpha_n $, $s^\alpha_s=1/k^\alpha_s$, and $s^\alpha_t=1/k^\alpha_t$ are the reciprocals of the local stiffness coefficients.

\subsection{Integral form of the formulation}

It bears emphasis that the relationships given in equations \eqref{Residual}-\eqref{epsilon-Sijkl} are in summation form over all pair-interactions within the granular system that constitutes the representative volume element of interest. However, both the inter-particle force-displacement relationships and geometrical properties, such as the relative distance between interacting particles, depend strongly on direction with respect to a reference frame. It is then convenient to derive an integral form of the constitutive relationship by defining two particle-scale  directional density distribution functions, namely the directional distribution of a combined mechano-geometrical property and the directional distribution of a purely geometrical property \cite{Payam-Gibbs}. In this formulation, for convenience, a global spherical coordinate system is utilized wherein $\theta$, $\phi$, and $\rho$ denote the polar zenith angle, the azimuth angle, and the radial coordinate, respectively (see Figure \ref{fig-nst}). Specifically, for defining the fabric tensor $N_{ij}$ in integral form, a directional distribution $\xi'(\theta,\phi)$ of the particle-particle relative distance squared, i.e., of $||\qvec{l}||^2$, is proposed as follows
\begin{equation}
\begin{aligned}
N_{ij}=&l^2\rho_c\int_{\theta=0}^{\pi}\int_{\phi=0}^{2\pi}n_i n_j \xi'\left(\theta,\phi\right) \sin\theta d\theta d\phi; 
\\
&\ \ \text{with} \ \ 
\xi'\left(\theta,\phi\right)
=
{\sum\limits_{\rho\in\mathcal{P}_{\theta\phi}} ||\qvec{l}^\rho||^2}/{\sum\limits_{\alpha=1}^{N_c} ||\qvec{l}^\alpha||^2}
\end{aligned} 
\label{Eq-Nij}
\end{equation}
where $l$, $\rho_c=N_c/V$ and $\mathcal{P}_{\theta\phi}$ denote the scalar values of average inter-particle length, the volume density of contacts in the assembly, and the subset of contacts in direction $(\theta,\phi)$, respectively. Similarly, for defining the local compliance tensor $s_{ij}$ in integral form, a directional distribution of combined mechano-geometrical properties $s_n||\qvec{l}||^2$,  $s_s||\qvec{l}||^2$ and  $s_t||\qvec{l}||^2$---all assumed the same for simplicity---is proposed as follows
\begin{equation}
\begin{aligned}
\xi\left(\theta,\phi\right)
&=
{\sum\limits_{\rho\in\mathcal{P}_{\theta\phi}} 
\left( s_n^\rho||\qvec{l}^\rho||^2\right)}
/
{\sum\limits_{\alpha=1}^{N_c} \left(s_n^\alpha||\qvec{l}^\alpha||^2\right)}
\\ 
&=
{\sum\limits_{\rho\in\mathcal{P}_{\theta\phi}} 
\left( s_s^\rho||\qvec{l}^\rho||^2\right)}
/
{\sum\limits_{\alpha=1}^{N_c} \left(s_s^\alpha||\qvec{l}^\alpha||^2\right)}
\\
&=
{\sum\limits_{\rho\in\mathcal{P}_{\theta\phi}} 
\left( s_t^\rho||\qvec{l}^\rho||^2\right)}
/
{\sum\limits_{\alpha=1}^{N_c} \left(s_t^\alpha||\qvec{l}^\alpha||^2\right)}
\end{aligned}
\label{Eq-xi-sl2}
\end{equation}
Therefore, the macroscopic compliance tensor in integral form simplifies to
\begin{equation}
S_{ijkl}=l^2\rho_cN_{jp}^{-1}N_{lq}^{-1}
\int_{\theta=0}^{\pi}\int_{\phi=0}^{2\pi}s_{ik} n_p n_q \xi\left(\theta,\phi\right) \sin\theta d\theta d\phi
\label{Eq-Sijkl}
\end{equation}
where $s_{ik}$ is the average local compliance tensor which can be expressed in local coordinates as $\qvec{s}=s_n \qvec{n}\otimes\qvec{n}+s_s \qvec{s}\otimes\qvec{s}+s_t \qvec{t}\otimes\qvec{t}$, with $s_n$, $s_s$ and $s_t$ denoting the reciprocal of the average local stiffness coefficients $k_n$, $k_s$ and $k_t$, respectively. Finally, spherical harmonics, in terms of Legendre polynomials, are utilized for defining these two directional probability distribution functions, i.e.,
\begin{equation}
	\begin{split}
	\xi(\theta,\phi)= & \frac{1}{4\pi}\Bigg\{1+\sum_{k=2}^{\infty}{}^{'} \Big[a_{k0}P_k\left(\cos\theta\right)+ \\
	& \sum_{m=1}^{\infty}P_k^m\left(\cos\theta\right)\big(a_{km}\cos m\phi+b_{km}\sin m\phi\big)\Big]\Bigg\}
	\end{split}
	\label{Eq-xi}
\end{equation}
where the summation over $k$, represented by $\sum'$, denotes summation over even values of $k$, $P_k\left(\cos\theta\right)$ is the $k^\text{th}$ order Legendre polynomial with respect to $\cos\theta$, while $P_k^m\left(\cos\theta\right)$ is its $m^\text{th}$ associated Legendre function. Parameters $a_{k0}$, $a_{km}$, and $b_{km}$ are fabric parameters governing the shape of the distribution function. Note that the relationship for $\xi'$ will be identical to that presented in Eq. \ref{Eq-xi}, but will use $a'_{k0}$, $a'_{km}$, and $b'_{km}$ as fabric parameters.  Since Legendre polynomials and their associated functions are all orthogonal to one, the integral of both $\xi$ and $\xi'$ over the surface of a unit sphere is identical to unity, regardless of the number of fabric parameters used and their representation. This property is consistent with the definition of the two directional probability distribution functions given in Eqs. \ref{Eq-Nij} and \ref{Eq-xi-sl2}.

\subsection{Transversely isotropic materials}

The mechanical response of isotropic materials is identical in all direction; therefore, the directional distribution of contact properties can be adopted  constant in all directions and thus the fabric parameters in $\xi$ and $\xi'$ are equal to 0. In addition, the inter-particle stiffness coefficients in the two tangential directions can be assumed equal to each other, i.e., $k_s=k_t$.

In contrast, the mechanical response of transversely isotropic materials is independent of $\phi$, but depends on $\theta$, and thus only $a_{k0}$ and $a'_{k0}$ are different from 0. The fabric tensor $N_{ij}$ for a transversely isotropic material derived from Eq. \eqref{Eq-Nij} using all even terms $a'_{k0}$ simplifies to
\begin{equation}
N_{ij}=\frac{l^2\rho_c}{15} \begin{bmatrix}
5-a'_{20} & 0 & 0\\ 0 & 5-a'_{20} & 0 \\ 0 & 0 & 5+2a'_{20}
\end{bmatrix}
\label{Nij-trans}
\end{equation}
where only the dependency on $a'_{20}$ emerges. Similarly, using Eqs. \eqref{Eq-Sijkl} and \eqref{Eq-xi} with all even terms $a_{k0}$, the compliance tensor $S_{ijkl}$ for a transversely isotropic material retains a dependency only on $a_{20}$ and $a_{40}$. It is worth noting then that the components of the macroscopic stiffness tensor $C_{ijkl}$, and of its inverse $S_{ijkl}$, are functions of only 5 parameters, i.e.,  
\begin{equation}
S_{ijkl}^{-1}
=
C_{ijkl}
=
C_{ijkl}\left(l^2\rho_c k_n,l^2\rho_c k_s, a_{20},a_{40},a'_{20}\right)
\end{equation}
where $k_s=k_t$ is assumed. Specifically, the five independent components of the compliance tensor of a transversely isotropic material can be derived in closed-form as follows
\begin{equation}
\begin{aligned}
& S_{11}=\frac{15}{7 (a'_{20}-5)^2}\left(  \frac{21 - 6 a_{20} + a_{40}}{k_n l^2\rho_c}  + \frac{14 - a_{20} - a_{40}}{k_s l^2\rho_c}  \right)\\
& S_{33}=\frac{5}{7 (2a'_{20}+5)^2}\left(  \frac{63 +36 a_{20} + 8a_{40}}{k_n l^2\rho_c}  + \frac{42 +6 a_{20} -8 a_{40}}{k_s l^2\rho_c}  \right)\\
& S_{44}=\frac{5}{7 (a'_{20}-5)^2(2a'_{20}+5)^2}\bigg(  (a'_{20}+10)^2\frac{21 +3 a_{20} - 4a_{40}}{k_n l^2\rho_c} +\\
& \frac{126(25+5a'_{20}+4a'^2_{20}) + 45(5 - 20a'_{20} - a'^2_{20}) + 4a_{40}(10 + a'_{20})^2}{k_s l^2\rho_c}  \bigg)\\
& S_{12}=\frac{5}{7 (a'_{20}-5)^2}\left( 21-6 a_{20} + a_{40} \right)\frac{k_s-k_n}{k_nk_s l^2\rho_c}\\
& S_{13}=\frac{-5}{7 (a'_{20}-5)(2a'_{20}+5)}\left( 21 + 3 a_{20} - 4a_{40} \right)\frac{k_s-k_n}{k_nk_s l^2\rho_c}\\
\end{aligned}
\label{Eq-Sij-Trans}
\end{equation}
These five independent components of the compliance tensor---with $S_{66}=2\left(S_{11}-S_{12}\right)$---are obtained from the fourth order compliance tensor by symmetrizing its components in the form of a $6\times6$ matrix \cite{Malvern-Continuum}. The macroscopic compliance tensor captures both the inherent and loading-induced anisotropies automatically and at a minimal increase in computational demand---cf. discrete particle mechanics simulations, e.g., \cite{Gonzalez-2018}. The trade-off between a modeling approach which is fully descriptive at the particle scale and a modeling approach which is descriptive in an average directional sense is favorable when the macroscopic behavior is of primary interest \cite{Weiner-StatisticalElasticity}. In this study, we seek to derive the evolution of average inter-particle stiffness coefficients, $k_n$ and $k_s$, as well as of the directional distribution of contact properties, characterized by fabric parameters $a_{20}$, $a_{40}$ and $a'_{20}$, during a conventional triaxial loading process such as die compaction of granular systems up to relative densities close to 1, or porosities close to 0. We propose to achieve this goal by connecting discrete particle mechanics to continuum granular micromechanics.

\begin{figure}[htbp]
    \centering{
	\begin{tabular}{ccc}
	\includegraphics[scale=.90]{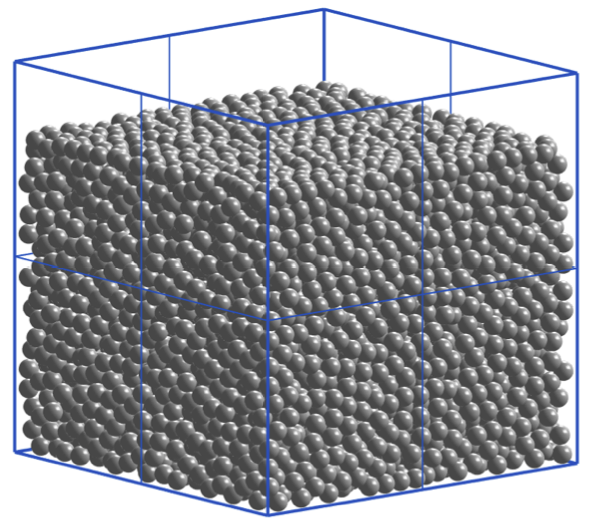}
	&
	\includegraphics[scale=.90]{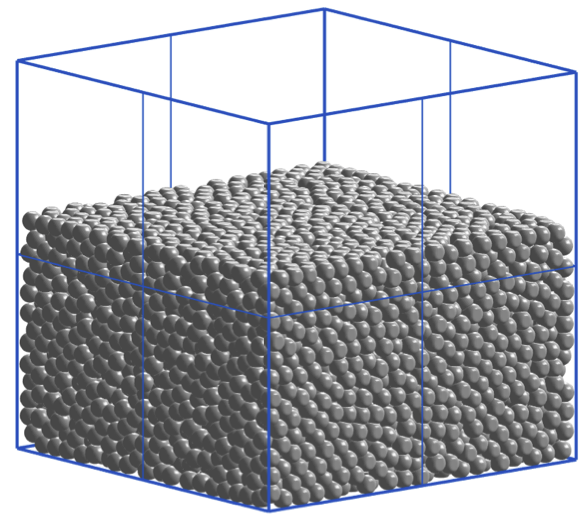}
	&
	\includegraphics[scale=.90]{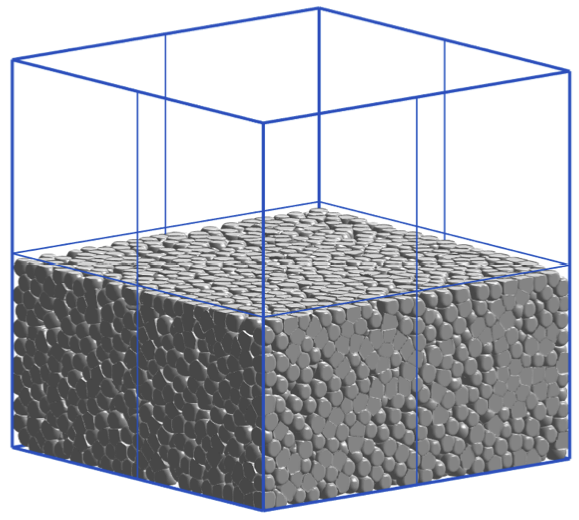}
	\end{tabular}	    
    }
	\caption{Deformed three-dimensional packing at three different stages: (left) unload state ($\epsilon_{33}=0$ and relative density $\rho=0.58$), (middle) intermediate stage ($\epsilon_{33}=26\%$ and relative density $\rho=0.89$), and (right) fully compaction stage ($\epsilon_{33}=41\%$ and relative density $\rho=0.99$).}
	\label{fig-packing}
\end{figure}

\section{Connecting Discrete to Continuum}
In this study, we restrict attention to a monodisperse packing of elastic spherical grains, located in a  box with a square base and loaded under die-compaction conditions, i.e., under triaxial loading. We adopt a particle mechanics approach (PMA) which describes each individual particle in the packing, and the collective rearrangement and deformation of the particles that result in a quasi-statically compacted specimen \cite{Gonzalez-2012,Gonzalez-2016,yohannes2016evolution}. The loading conditions consist of applying a vertical displacement to the upper wall of the confining box (i.e., applying a macroscopic $\epsilon_{33}$) while keeping the lateral walls unperturbed (i.e., enforcing a macroscopic  $\epsilon_{11}=\epsilon_{22}=0$). We use the Hertz contact theory for elastic spherical particles \cite{timoshenko1970theory}, and thus the contact force $F$ between two identical spherical particles with radius $R$, elastic modulus of $E$ and Poisson's ratio of $\nu$, is given by  
\begin{equation}
	F
	=
	n_\mathrm{H}(\gamma)_+^{3/2} 
	\ \ \ \text{where } \ \ \ 
	n_\mathrm{H}
	=
	E (2 R)^{1/2}/3(1-\nu^2)\\
\label{Hertz-elastic}
\end{equation}
where $( \cdot )_+ = \max\{\cdot, 0\}$. We specifically study a noncohesive frictionless granular system comprised of $7,357$ weightless spherical particles with radius $R=250\mu$m and elastic properties $E=7$GPa, $\nu=0.30$ (see Figure \ref{fig-packing}). Due to the elastic nature of the particles, at a given macroscopic strain $\qvec{\epsilon}^t$, the total internal strain energy density of the system is then given by the volume average of the inter-particle energies, for which a close-form solution is attainable
$$
	W^t
	= 
	\frac{1}{V_t}\sum_{\alpha=1}^{N_c}
	\left(\int F^\alpha\mathrm{d}\gamma^\alpha\right)
	= 
	\frac{1}{V_t}\sum_{\alpha=1}^{N_c}
	\frac{2}{5} n_\mathrm{H}(\gamma^\alpha)_+^{5/2} 
$$
where $V_t$ is the packing volume in the current configuration $t$, while $F^\alpha$ and $\gamma^\alpha$ represent the inter-particle force and displacements. 

\begin{figure}[htbp]
	\centering{
	\includegraphics[scale=0.620]{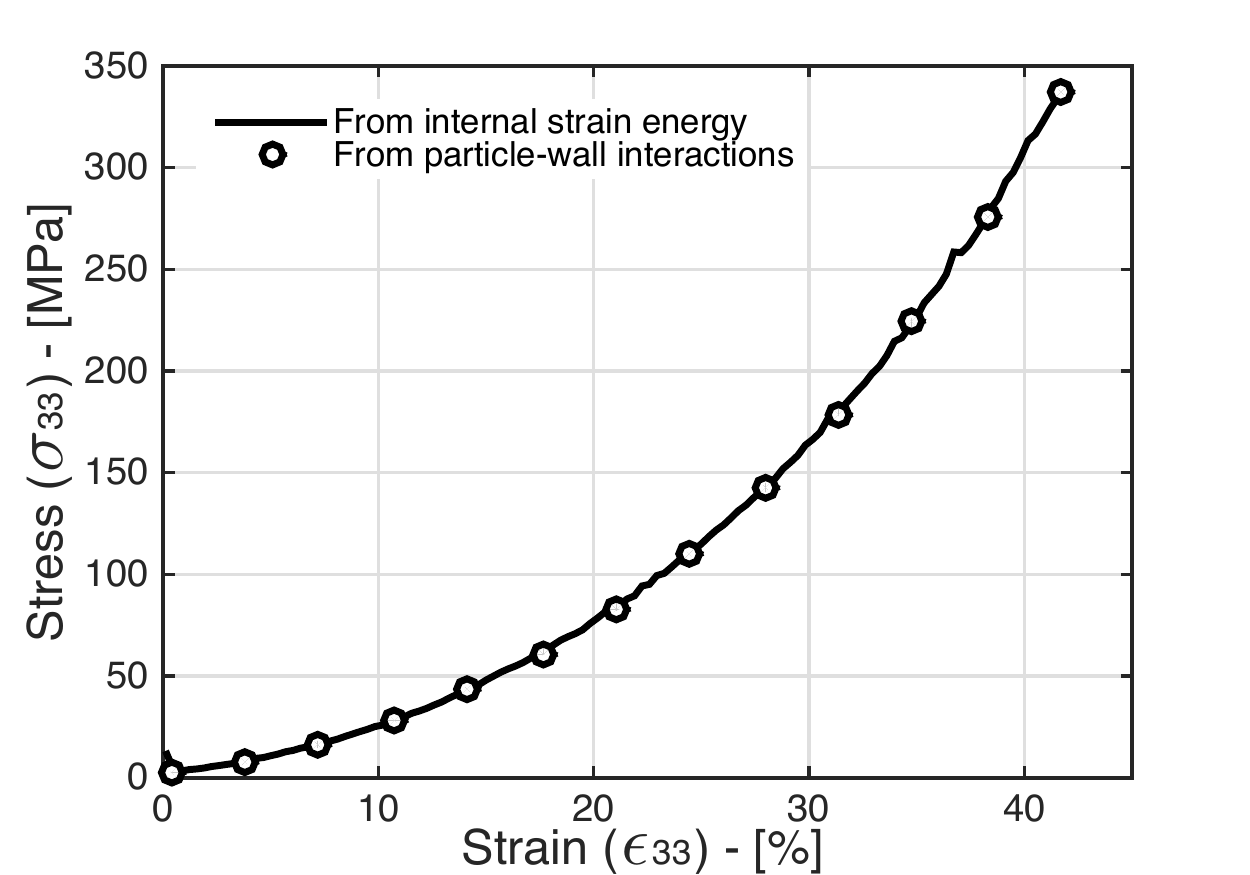}
	}
	\caption{Stress-strain curves determined from the evolution of the internal strain energy density and from particle-wall interactions of particle mechanics simulations.}
	\label{fig-stress}
\end{figure}

Figure \ref{fig-stress} shows the stress in the direction of compaction, $\sigma_{33}$, determined from the particle-wall interactions of PMA simulations (i.e., the sum of particle-wall forces divided by the wall area), and from the evolution of the total internal strain energy density in the packing, i.e.,
\begin{equation}
	\sigma_{33}^t =
	\frac{\mathrm{d}W}{\mathrm{d}\epsilon_{33}}
	=
	\frac{W^{t'} - W^t }{\mathrm{d}\epsilon_{33}}
	\label{w-sigma-eps}
\end{equation}
where $W^{t'}$ corresponds to the strain energy density of the packing perturbed by $\mathrm{d}\epsilon_{33}$. It is evident from the figure that there is perfect agreement between the two results, which confirms the correct definition of work conjugates.

Due to the symmetries in the initial random packing, as well as the symmetry in the applied load during compaction, it is assumed that the discrete packing behaves as a transversely isotropic material. Next, we focus on deriving, from PMA simulations, the evolution of all five independent components of the stiffness tensor $C_{ij}$ during die compaction and, subsequently, we identify the evolution of five GMA parameters which describe the same macroscopic material behavior. In order to calculate the components of the stiffness tensor, at every loading step $t$ during compaction, perturbations in different strain components, $\mathrm{d}\epsilon_{ij}$, are applied using the PMA and the change of stored strain energy density, $\mathrm{d}W$, is calculated. Furthermore, for every perturbation at loading step $t$, $\mathrm{d}W$ can be written in terms of the tangent stiffness tensor $C_{ijkl}^t$ as
\begin{equation}
\mathrm{d}W = \sigma_{ij}^t \mathrm{d}\epsilon_{ij} + \frac{1}{2}\mathrm{d}\sigma_{ij}\mathrm{d}\epsilon_{ij} = \sigma_{ij}^t \mathrm{d}\epsilon_{ij} +
 \frac{1}{2} \mathrm{d}\epsilon_{ij}  C_{ijkl}^t \mathrm{d}\epsilon_{kl} 
\label{C-from-W}
\end{equation}

\begin{figure}[tbp]
	\centering
	\includegraphics[scale=0.620]{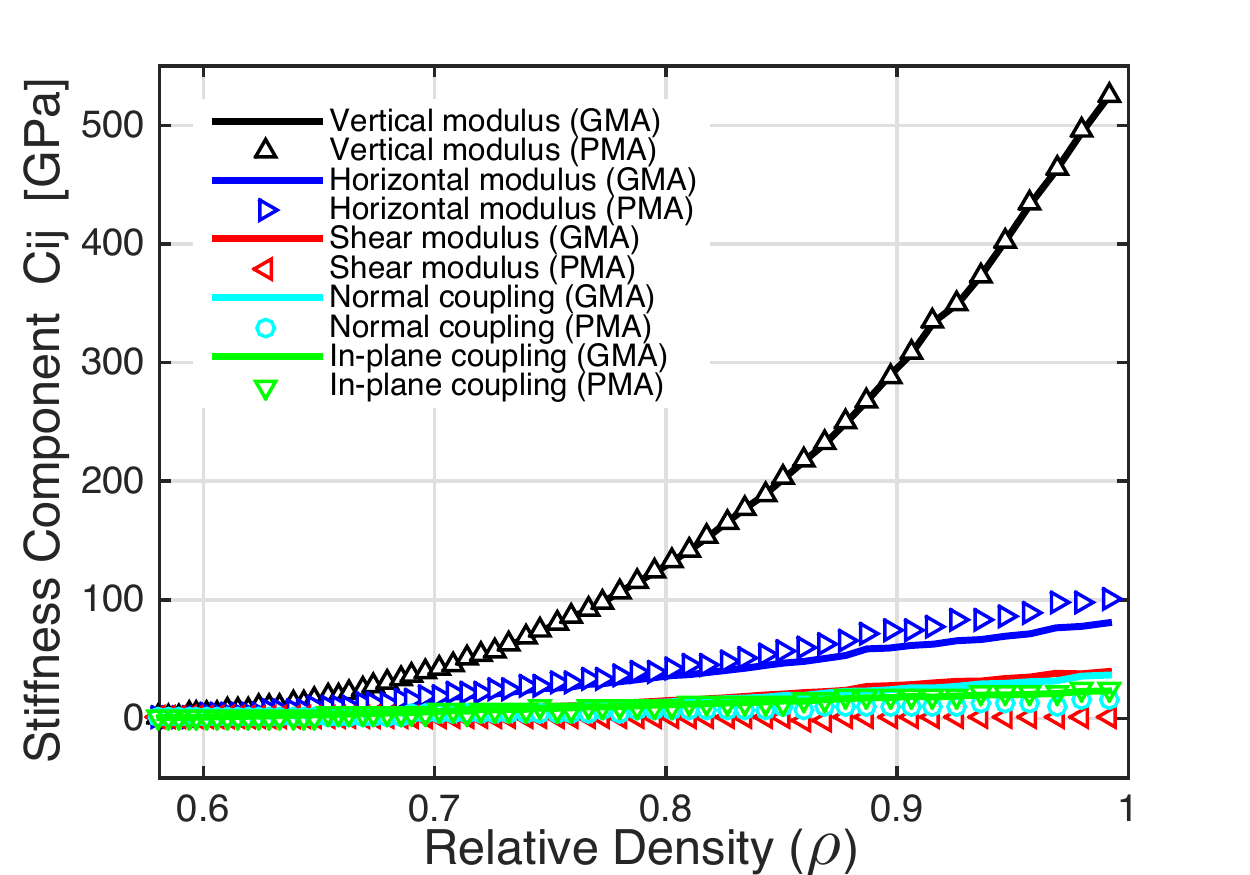}
	\caption{Evolution of stiffness components during compaction. Numerical results obtained from the PMA (solid lines) compared to GMA results using the set of 5 best-fitted parameters (symbols)}
	\label{fig-CijFit}
\end{figure}

In order to calculate all elements of the stiffness tensor, 9 combinations of small perturbations in the strain tensor are applied to the packing. These perturbations include applying six uni-directional strains, namely $\mathrm{d}\epsilon_{11}$, $\mathrm{d}\epsilon_{22}$, $\mathrm{d}\epsilon_{33}$, $\mathrm{d}\epsilon_{23}$, $\mathrm{d}\epsilon_{13}$, and $\mathrm{d}\epsilon_{12}$, and three combinations of $\mathrm{d}\epsilon_{11}$ and $\mathrm{d}\epsilon_{22}$, $\mathrm{d}\epsilon_{11}$ and $\mathrm{d}\epsilon_{33}$, and $\mathrm{d}\epsilon_{22}$ and $\mathrm{d}\epsilon_{33}$. After calculating the variation of strain energy density caused by these perturbations, all 9 nonzero components of the stiffness tensor, namely three normal diagonal components ($C_{1111}$, $C_{2222}$ and $C_{3333}$), three shear components ($C_{2323}$, $C_{1313}$ and $C_{1212}$), and three coupling components ($C_{1122}$, $C_{1133}$ and $C_{2233}$) are determined by solving a linear system of 9 equations and 9 unknowns. Figure \ref{fig-CijFit} shows the evolution of symmetrized stiffness components calculated using  the system of equations derived by applying the above perturbations and solving Eq. \eqref{C-from-W}. It is important to point out that the three coupling components of stiffness tensor, i.e., the in-plane coupling $C_{1122}$ and the coupling between vertical and in-plane directions $C_{1133}$ and $C_{2233}$, are almost equal. Moreover, it is seen that the shear stiffness of the packing remains very close to zero throughout compaction. 

Finally, the GMA parameters needed for modeling transversely isotropic materials, i.e., $l^2\rho_c k_n$, $l^2\rho_c k_s$, $a_{20}$, $a_{40}$, and $a'_{20}$, are calculated by solving an optimization problem aimed at best fitting the components of the macroscopic stiffness tensor. Closed-form solutions for components of the stiffness tensor are obtained by inverting the closed-form solution of the compliance tensor given in Eq. \eqref{Eq-Sij-Trans}. Figure \ref{fig-CijFit} shows the evolution of stiffness components obtained by using the best-fitted GMA parameters. It is worth noting that the optimization problem is subjected to the inequality constraint of positive $\xi(a_{20},a_{40})$ and $\xi'(a'_{20})$ in all directions and, naturally, of positive $l^2\rho_c k_n$ and $l^2\rho_c k_s$. It is evident from the figure that the GMA analysis accurately predicts the evolution of the macroscopic stiffness components from micromechanical parameters, even up to full compaction (or relative density $\rho=1$) where finite macroscopic deformations occur. However, there is some discrepancy in the fitting of the shear modulus and a more clear investigation of packing size effects, boundary effects, and GMA assumptions is desirable, if beyond the scope of this study.

\begin{figure}[htbp]
	\centering{
	\begin{tabular}{cc}
	\includegraphics[scale=0.620]{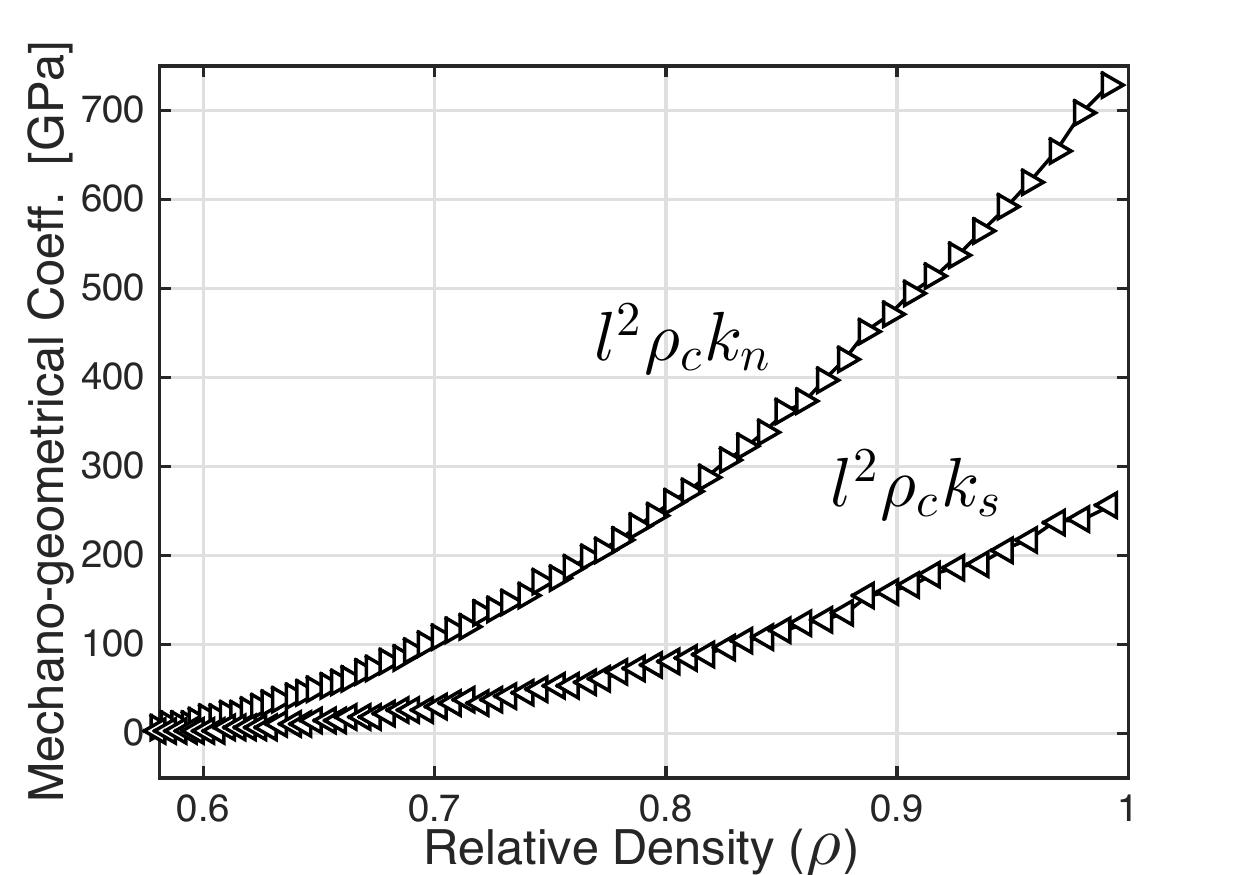}
	\includegraphics[scale=0.620]{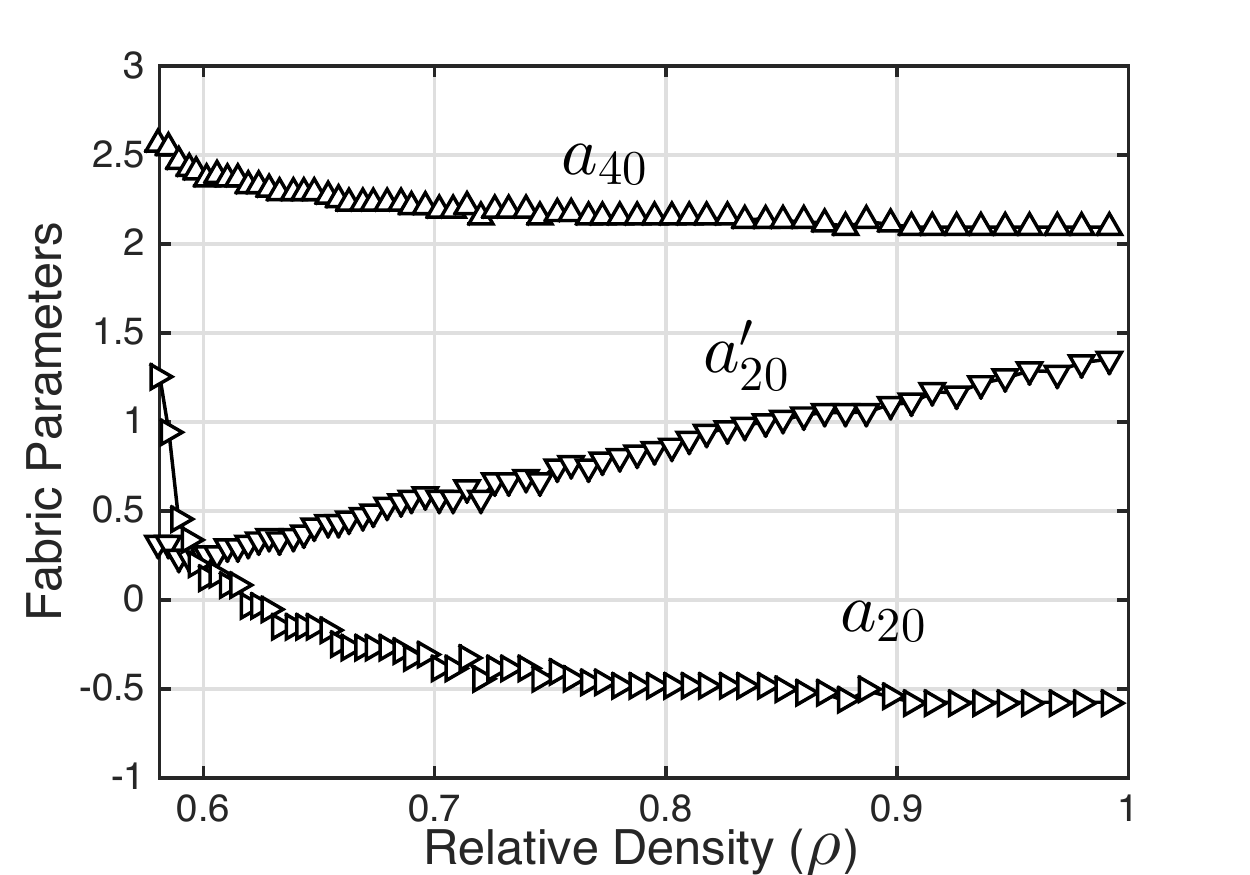}	
	\end{tabular}
	}
	\caption{Evolution of the optimized GMA parameters during compaction. Left: mechano-geometrical coefficients  $l^2\rho_c k_n$ and $l^2\rho_c k_s$. Right: fabric parameters $a_{20}$, $a_{40}$ and $a'_{20}$.}
	\label{fig-GMAparam}
\end{figure}

\begin{figure}[htbp]
    \centering{
	\begin{tabular}{cc}
	\includegraphics[trim= 29mm 10mm 25mm 5mm, clip, scale=0.59]{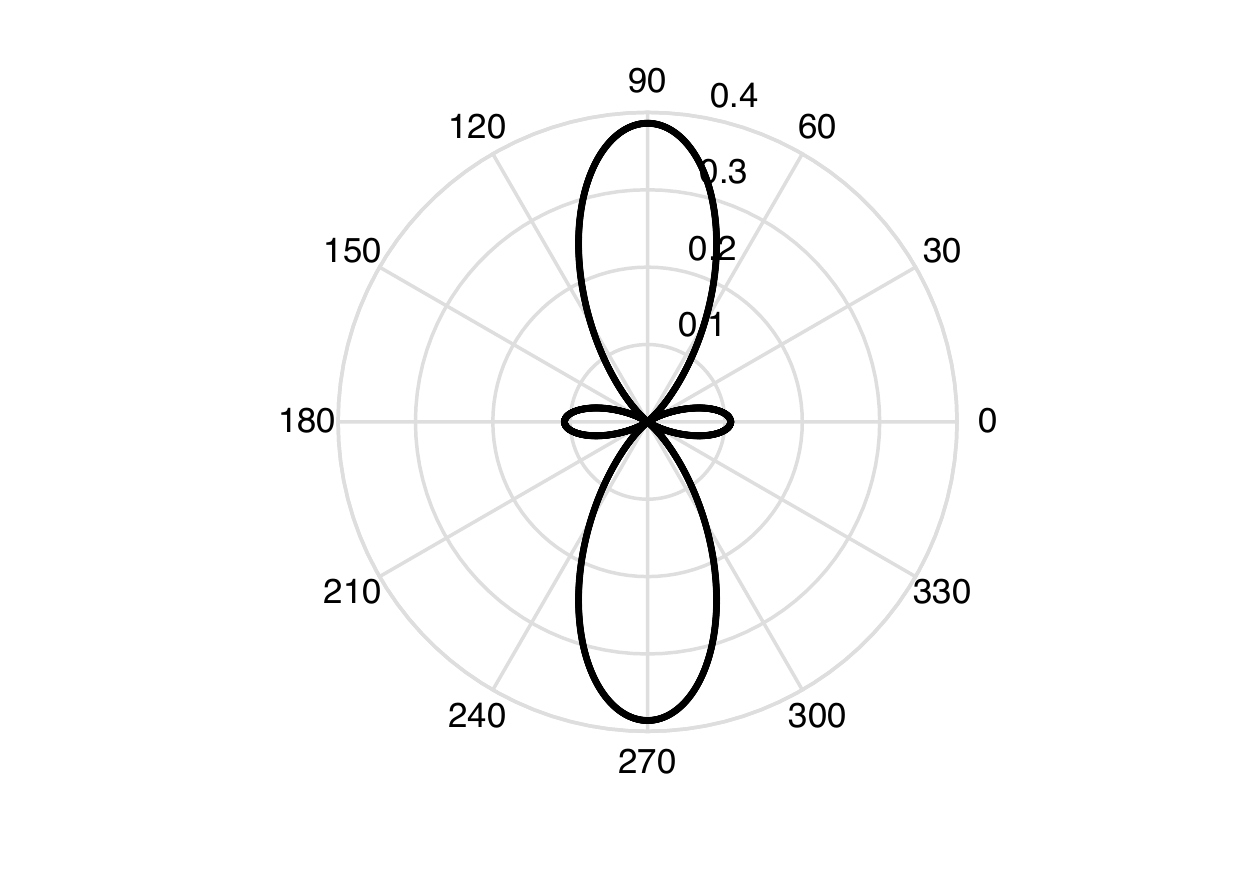}
	&
	\includegraphics[trim= 29mm 10mm 25mm 5mm, clip, scale=0.59]{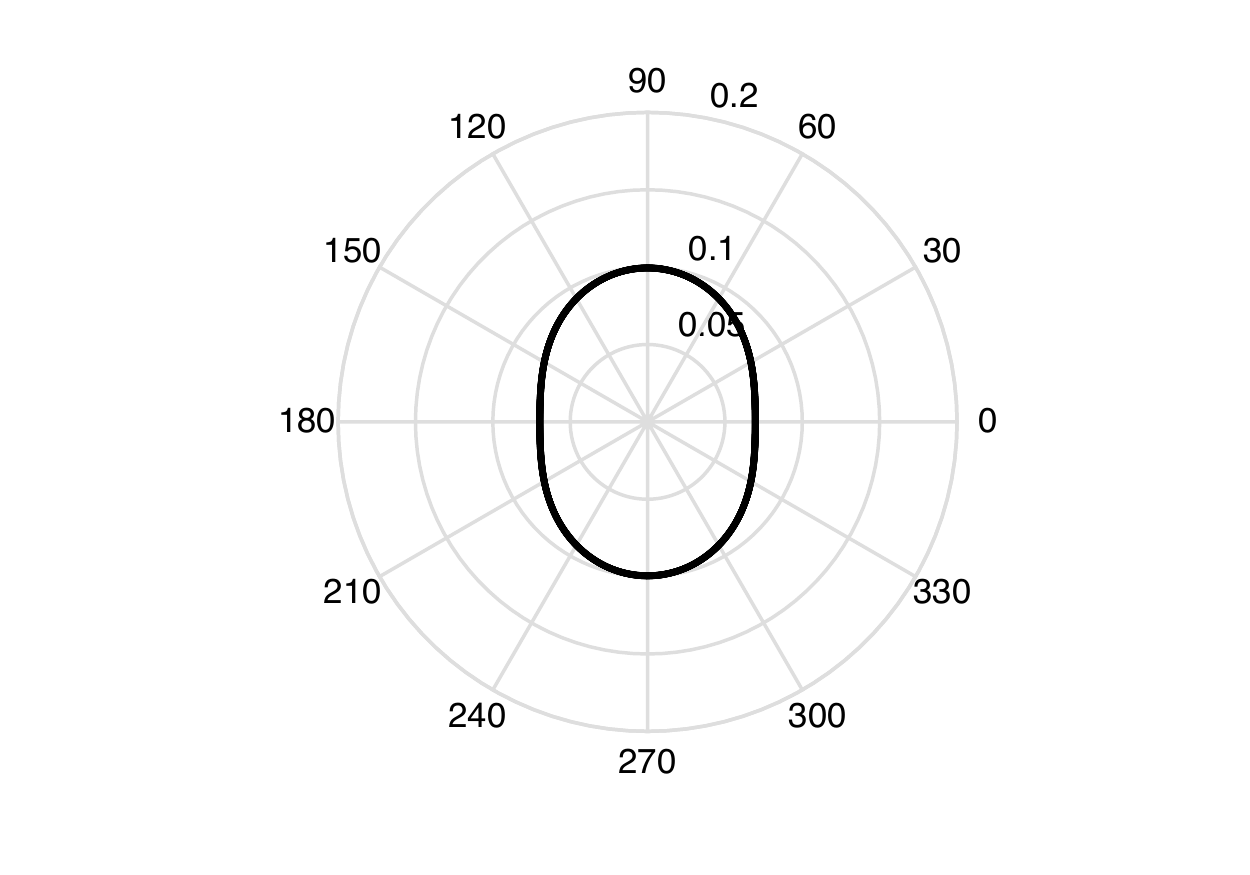}
	\\
	\includegraphics[trim= 29mm 10mm 25mm 5mm, clip, scale=0.59]{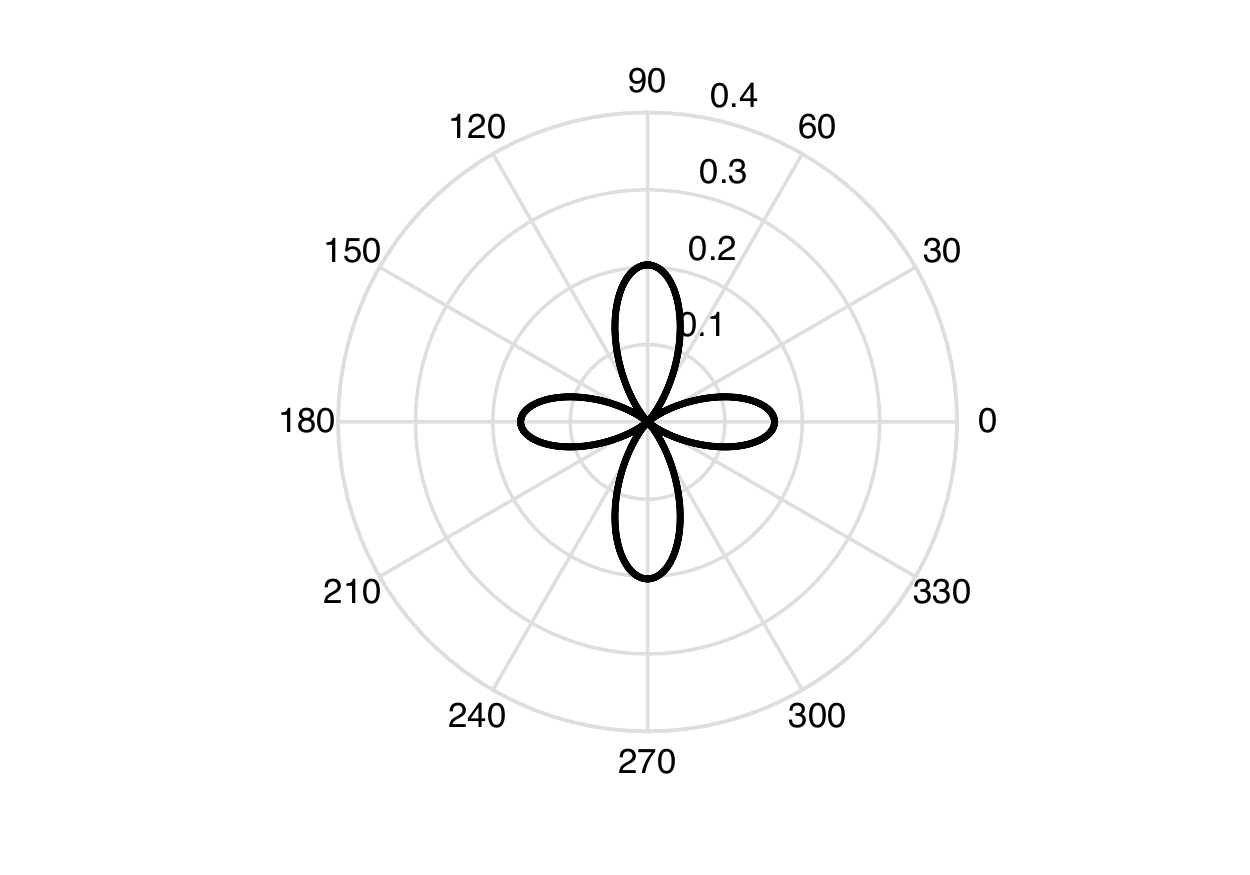}
	&
	\includegraphics[trim= 29mm 10mm 25mm 5mm, clip, scale=0.59]{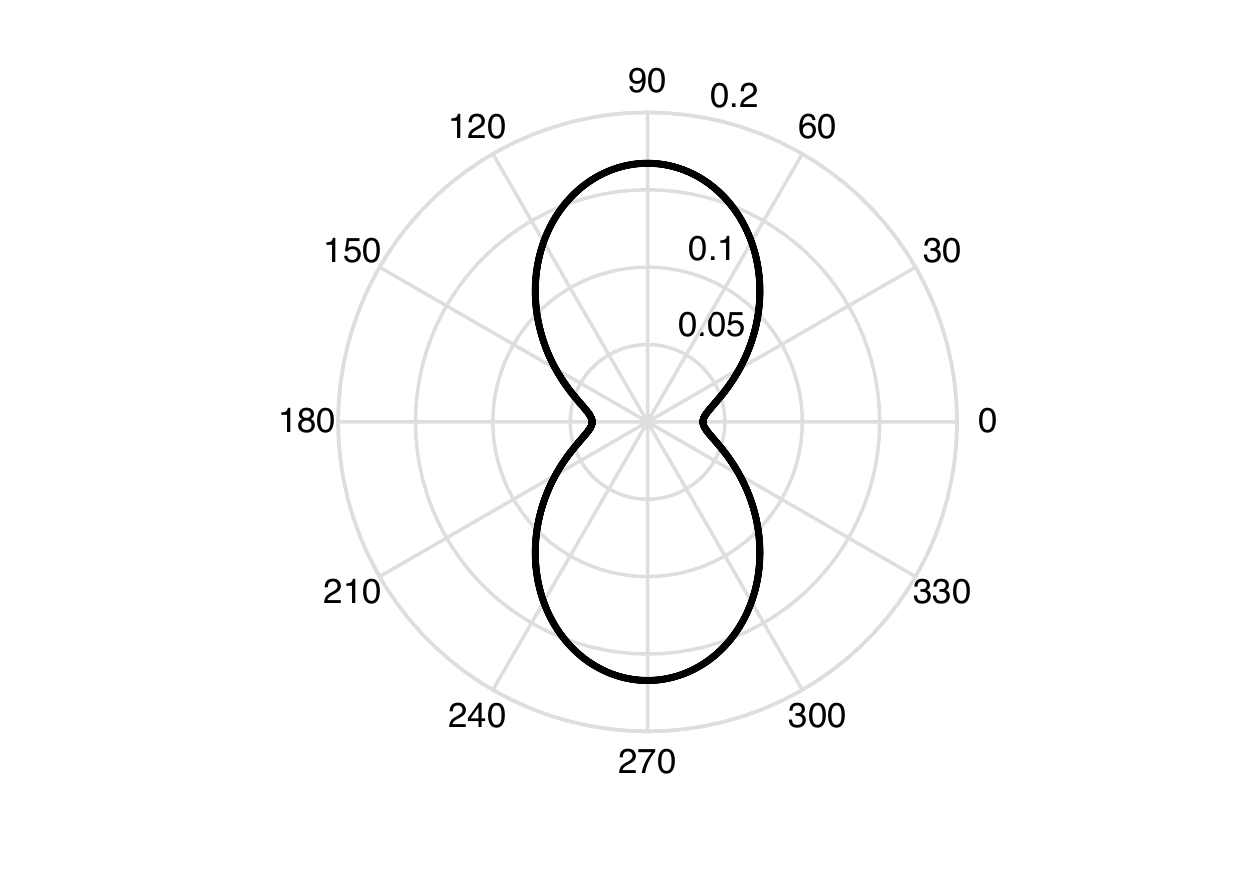}
	\\
	\includegraphics[trim= 29mm 10mm 25mm 5mm, clip, scale=0.59]{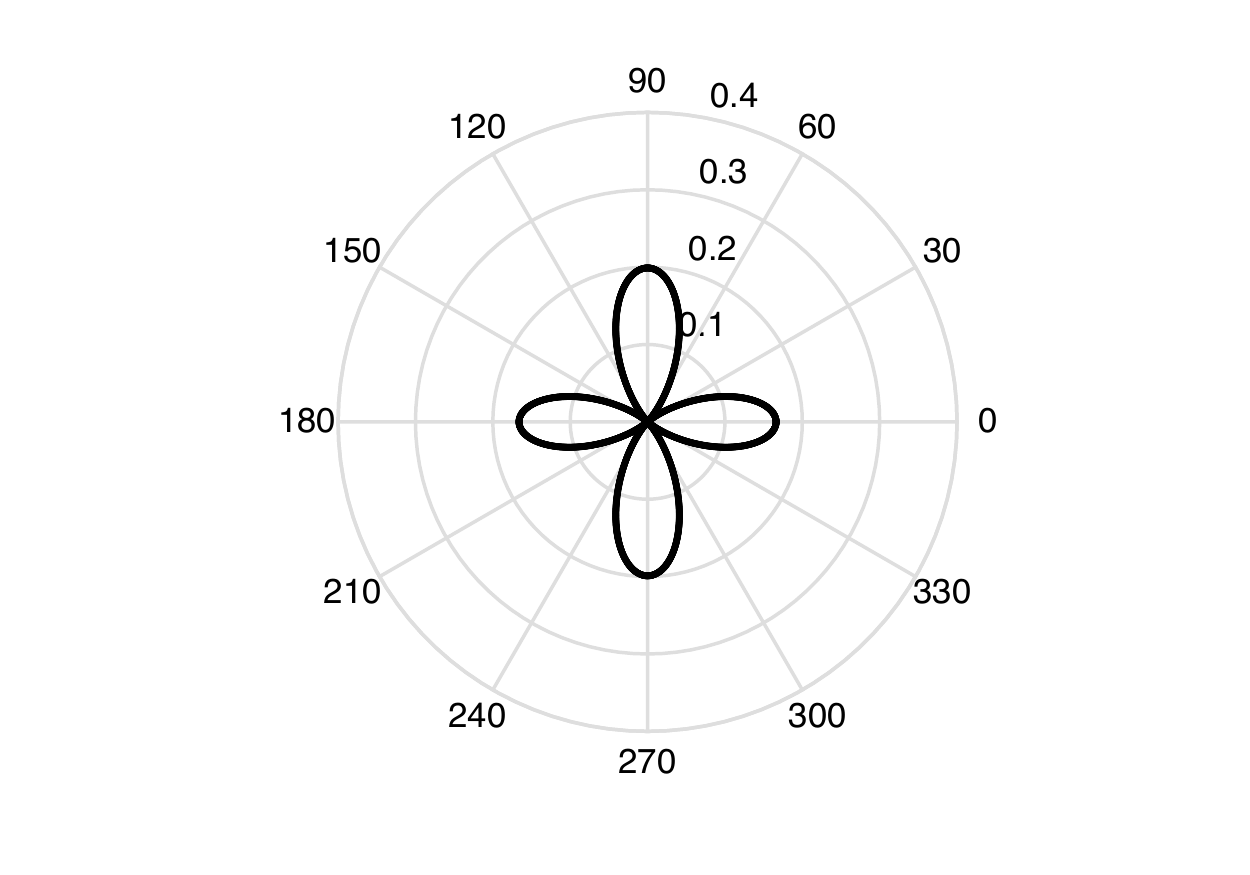}
	&
	\includegraphics[trim= 29mm 10mm 25mm 5mm, clip, scale=0.59]{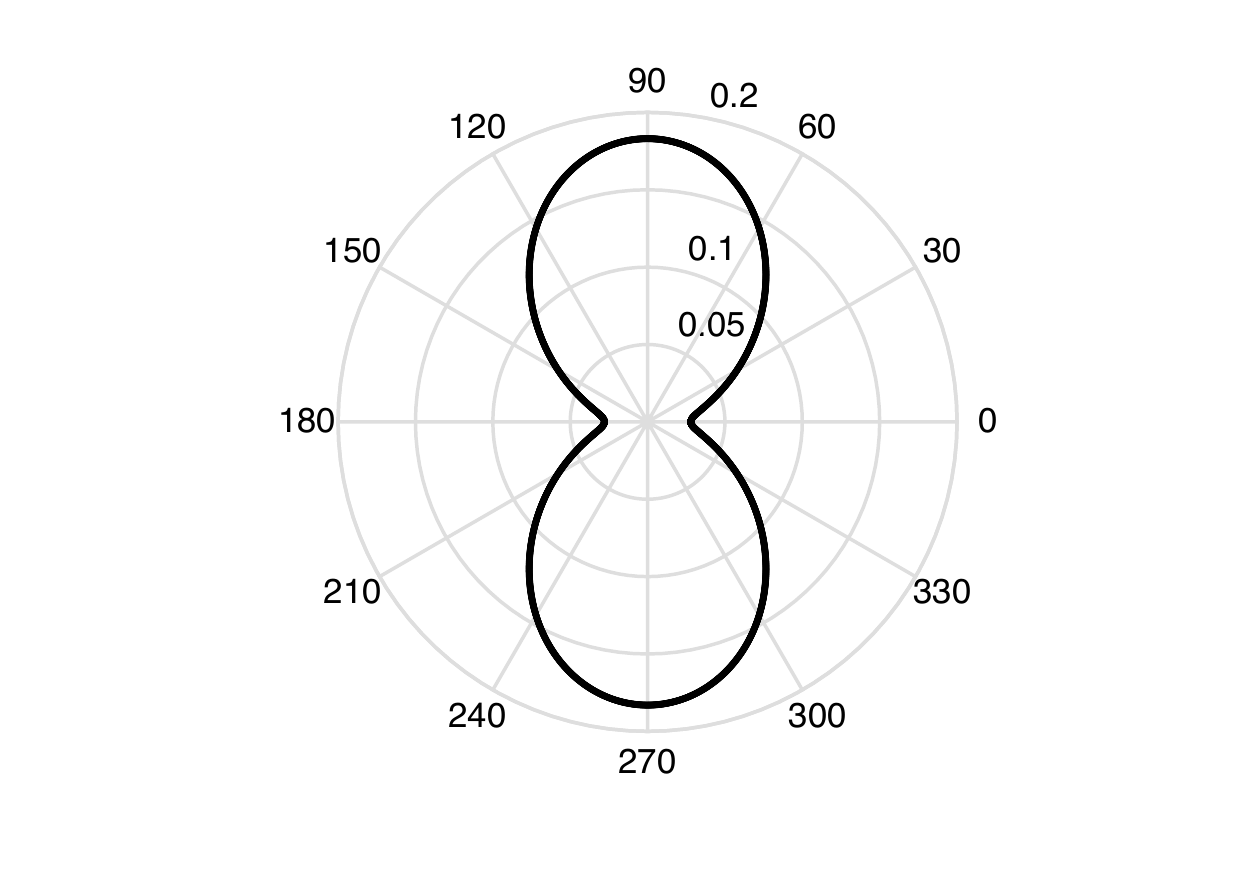}	
	\end{tabular}	    
    }
	\caption{Directional probability distribution functions $\xi$ (right) and $\xi'$ (left) at three different stages of compaction: (top) unload state ($\epsilon_{33}=0$ and relative density $\rho=0.58$), (middle) intermediate stage ($\epsilon_{33}=26\%$ and relative density $\rho=0.89$), and (bottom) fully compaction stage ($\epsilon_{33}=41\%$ and relative density $\rho=0.99$).}
	\label{fig-directional-distribution}
\end{figure}

Evolution of the optimized GMA parameters during compaction is shown in Figure \ref{fig-GMAparam}. The mechano-geometrical coefficients  $l^2\rho_c k_n$ and $l^2\rho_c k_s$, i.e., the product of contacts volume density $\rho_c$, average contact length squared $l^2$ and stiffness coefficients $k_n$ and $k_s$, show a smooth and monotonically ascending behavior, consistent with the stiffening of the granular material seen in Figure \ref{fig-CijFit}. The fabric parameters $a_{20}$ and $a_{40}$ also show monotonic behavior but exhibit a clear asymptotic trend towards $-0.50$ and $2.00$, respectively; whereas the $a'_{20}$ shows a monotonically increasing behavior. Figure \ref{fig-directional-distribution} shows that these values for the fabric parameters ensure that the directional distribution functions are positive in all directions. The figure also shows that the directional probability distribution of $l^2\rho_c/k_n$ and $l^2\rho_c/k_s$, i.e., $\xi\left(\theta,\phi\right)$, has its minimum at $45^\circ$ which indicates that stronger contacts are formed at directions diagonal to the direction of compaction---which is consistent with the three-dimensional deformed packing depicted in Figure \ref{fig-packing}.

\section{Concluding Remarks}

The connection between the discrete behavior of a three-dimensional packing of elastic, spherical particles and its granular continuum behavior under quasi-static die-compaction was studied. Discrete PMA simulations, with Hertzian contact interactions, were used to inform a GMA continuum model that incorporates the directional distribution of inter-particle contact properties in an average sense. Specifically, the GMA with static assumption was effectively used in this work. A closed-form solution for the stiffness tensor $C_{ijkl}$ of transversely isotropic materials was derived using 5 GMA parameters, namely two mechano-geometrical coefficients $l^2\rho_c k_n$ and $l^2\rho_c k_s$, and three fabric parameters associated to two independent directional density distribution functions $\xi(a_{20},a_{40})$ and $\xi'(a'_{20})$---as theoretically demonstrated in \cite{Payam-Gibbs}. In addition, a robust methodology for computing 5 independent components of the tangent stiffness tensor of the transversely isotropic granular assembly using PMA calculations of perturbations around a loading step of interest was presented. Optimal values for the 5 GMA parameters were obtained by minimizing the error between PMA calculations and GMA predictions of stiffness tensor during compaction process. The results show that GMA with static assumption is capable of effectively capturing the anisotropic evolution of microstructure during loading, even without describing contacts independently but rather accounting for them in an average sense.

It is worth noting that these results suggest a one-to-one mapping between the 5 independent stiffness components of transversely isotropic materials with 5 independent GMA parameters which describe directional distribution of mechano-geometrical and of solely geometrical parameters. The functional dependency of these 5 GMA parameters with loading-path and macroscopic state variables remains elusive, but is of paramount importance for applying the GMA, informed by true mechanical and morphological properties of the granular system, to other general loading conditions. We believe the methodology presented in this work paves the way for addressing this challenge and, in turn, for exploring a large range of applications in various fields, including concrete \cite{VanMier,wriggers2006mesoscale,maekawa2003multi}, pharmaceutical  \cite{razavi2015general,razavi2018quantification,yohannes2015role}, dentin adhesive \cite{park2011influence}, explosive and energetic \cite{skidmore1998evolution,jordan2014mechanics} materials, which may favor from both detailed particle-scale mechanistic models but also from reduced-order models for manufacturing purposes \cite{Su-2018}.

\section*{Acknowledgments}
The authors gratefully acknowledge the support received from the National Science Foundation grant number CMMI-1538861 and from Purdue University's startup funds. M.G. also acknowledges the U.S. Air Force Office of Scientific Research for support through Award No. FA9550-15-1-0102 and the project’s program managers Dr. Martin Schmidt and Dr. Jennifer Jordan.

\bibliographystyle{IEEEtran}
\bibliography{IEEEabrv,MIMO}
\end{document}